\documentclass[lettersize,journal]{IEEEtran}
\usepackage{amsmath,amsfonts}
\usepackage{physics,amsmath}
\usepackage{algorithm}
\usepackage{algpseudocode}
\usepackage{multirow}
\usepackage{array}
\usepackage{bm}
\usepackage{subcaption}
\usepackage{mathtools}
\usepackage{xcolor}

\usepackage{textcomp}
\usepackage{stfloats}
\usepackage{lipsum}  
\usepackage{verbatim}
\usepackage{graphicx}
\usepackage{cite}
\usepackage{pdfpages}
\usepackage{tikz}
\usepackage{lettrine}
\usetikzlibrary{3d}
\usetikzlibrary{arrows.meta} % for more arrow tip kinds

\usepackage[colorlinks=true, linkcolor=blue, urlcolor=blue, citecolor=blue, anchorcolor=blue]{hyperref}

\begin{document}

\title{
DebriSense: Terahertz-based Integrated Sensing and Communications (ISAC) for Debris Detection and Classification in the Internet of Space (IoS)}
%A Novel Detection Approach for Molecular Communication: Leveraging Ligand-Receptor Interactions in the Frequency Domain?}
\author{Haofan Dong,~\IEEEmembership{Student Member,~IEEE},
        and Ozgur B. Akan,~\IEEEmembership{Fellow,~IEEE}
\thanks{The authors are with Internet of Everything Group, Department of Engineering, University of Cambridge, CB3 0FA Cambridge, UK.}% <-this % stops a space
\thanks{Ozgur B. Akan is also with the Center for neXt-generation Communications
(CXC), Department of Electrical and Electronics Engineering, Koç University, 34450 Istanbul, Turkey (email:oba21@cam.ac.uk)}}
	   
% \thanks{A preliminary version of this work was presented at IEEE International Conference on Communications (ICC) on June 2023.}
% <-this % stops a space



\maketitle

\begin{abstract}
The proliferation of Low Earth Orbit (LEO) satellite constellations has intensified the challenge of space debris management. This paper introduces DebriSense-THz, a novel Terahertz-Enabled Debris Sensing system for LEO satellites that leverages Integrated Sensing and Communications (ISAC) technology. We present a comprehensive THz channel model for LEO environments, incorporating debris interactions such as reflection, scattering, and diffraction. The DebriSense-THz architecture employs machine learning techniques for debris detection and classification using Channel State Information (CSI) features. Performance evaluation across different frequencies (30 GHz-5 THz), MIMO configurations, debris densities, and SNR levels demonstrates significant improvements in debris detection and classification accuracy (95-99\% at 5 THz compared to 62-81\% at 30 GHz). Higher SNR configurations enhance sensing performance, particularly at higher frequencies. The system shows robust performance across various debris densities and MIMO size in the THz range, with a noted trade-off between communication reliability and sensing accuracy at lower frequencies. DebriSense-THz represents a significant advance in space situational awareness, paving the way for more effective debris mitigation strategies in increasingly congested LEO environments.
\end{abstract}

\begin{IEEEkeywords}
ISAC, THz, space debris detection, space debris classification.
\end{IEEEkeywords}

\section{Introduction}

\lettrine{T}{he} proliferation of Low Earth Orbit (LEO) satellite constellations, exemplified by programs like StarLink and OneWeb, has been pivotal in advancing next-generation communication systems \cite{dominguez2020space, ucs_satellite_database}. However, the increasing accumulation of space debris poses a significant threat to the safety and longevity of these satellites. With an estimated 1 million debris pieces larger than 1 cm in diameter \cite{massimi2024deep} and countless smaller, untraceable objects \cite{nunez2015improving}, the need for accurate debris detection and tracking has never been more critical. The high-velocity nature of space debris \cite{miyamoto2019space} exacerbates the risk of catastrophic collisions, potentially triggering a cascade of debris creation known as the Kessler syndrome \cite{su1985contribution}.

Traditional space debris detection relies on Space Situational Awareness (SSA) programs, particularly Space Surveillance and Tracking (SST) systems \cite{kennewell2013overview}. These systems employ ground-based radar and optical detection methods \cite{vallado2011simulating,weeden2010global,muntoni2021crowded}. While ground-based optical detection excels at observing larger debris in higher orbits \cite{takeichi2021tethered}, both radar and optical systems struggle with detecting small, fast-moving debris \cite{montisci2019compact}.

Recent advancements have explored satellite-based observation methods and machine learning techniques to enhance debris detection capabilities \cite{liu2022space,tao2019deep,nunez2015improving,mahendrakar2021real,xi2020space}. These approaches, including deep learning \cite{tao2019deep} and reinforcement learning \cite{oakes2022double,siew2022optimal}, have shown promise in improving detection accuracy and efficiency. However, they often require dedicated hardware or substantial computational resources, limiting their scalability and integration with existing satellite systems.

Integrated Sensing and Communication (ISAC) represents a paradigm shift in wireless technology, combining sensing and communication functionalities through shared resources \cite{yin2024integrated,liu2022integrated}. In the context of space debris detection, ISAC leverages Terahertz (THz) band properties, offering significant advantages for both communication and detection \cite{marchetti2022space}. Although current LEO satellites primarily use lower frequency bands \cite{kumar2022snr}, the transition to THz is expected to support communication rates up to 1 Tbps \cite{song2023analysis}, allowing highly directional beams and dense antenna arrays \cite{kumar2022snr}. Despite propagation challenges, THz technologies are well-suited for space applications \cite{imt2030_2021,imt2030_2022}, presenting a promising approach to enhance communication and debris detection in LEO networks.

In this paper, we introduce DebriSense-THz, a novel Terahertz-Enabled Debris Sensing system for LEO Satellites. To the best of our knowledge, this is the first study to use THz Integrated Sensing and Communications (TISAC) technology for the detection and classification of space debris in LEO satellite networks. Our technique exploits the unique properties of TISAC signals to simultaneously enhance communication capabilities and provide high-accuracy sensing for space debris detection and classification. The proposed DebriSense-THz system addresses critical limitations in current Space Situational Awareness (SSA) capabilities through the following key contributions:

\begin{enumerate}
\item Development of a comprehensive THz channel model for LEO satellite networks, incorporating debris interactions including reflection, scattering, and diffraction.
\item Design of a novel THz ISAC system architecture optimized for space debris detection and classification in LEO environments.
\item Implementation of an efficient machine learning approach using Support Vector Machines (SVM) for debris detection and classification based on Channel State Information (CSI) features.
\item Extensive performance evaluation of the DebriSense-THz system across various THz frequencies (0.3-5 THz), MIMO configurations, space debris densities ,and SNR levels.
\end{enumerate}

The remainder of this paper is organised as follows: Section II presents the architecture and methodology of the DebriSense-THz system, including the signal processing pipeline and the machine learning approach. Section III details a comprehensive multi-ray THz band signal channel model for LEO satellite communications in the presence of space debris. Section IV describes the simulation setup and presents a thorough performance evaluation of the DebriSense-THz system. Section V discusses system performance trade-offs and future directions, exploring adaptive design considerations and operational implications. Finally, Section VI concludes the paper, summarising key findings and their significance for space debris sensing in LEO environments.

\section{DebriSense-THz: System Architecture and Methodology}
The DebriSense-THz system takes advantage of the unique properties of TISAC to detect and classify space debris in LEO satellite networks. This section presents the overall system architecture, detailing the signal processing pipeline, feature extraction techniques, and the Support Vector Machine (SVM) based approach for debris detection and classification. The proposed system exploits the high sensitivity of THz signals to material properties, enabling accurate discrimination between different types of space debris while maintaining communication functionality.

\subsection{System Overview}

\begin{figure}[htbp]
  \centering
  \includegraphics[width=0.5\textwidth]{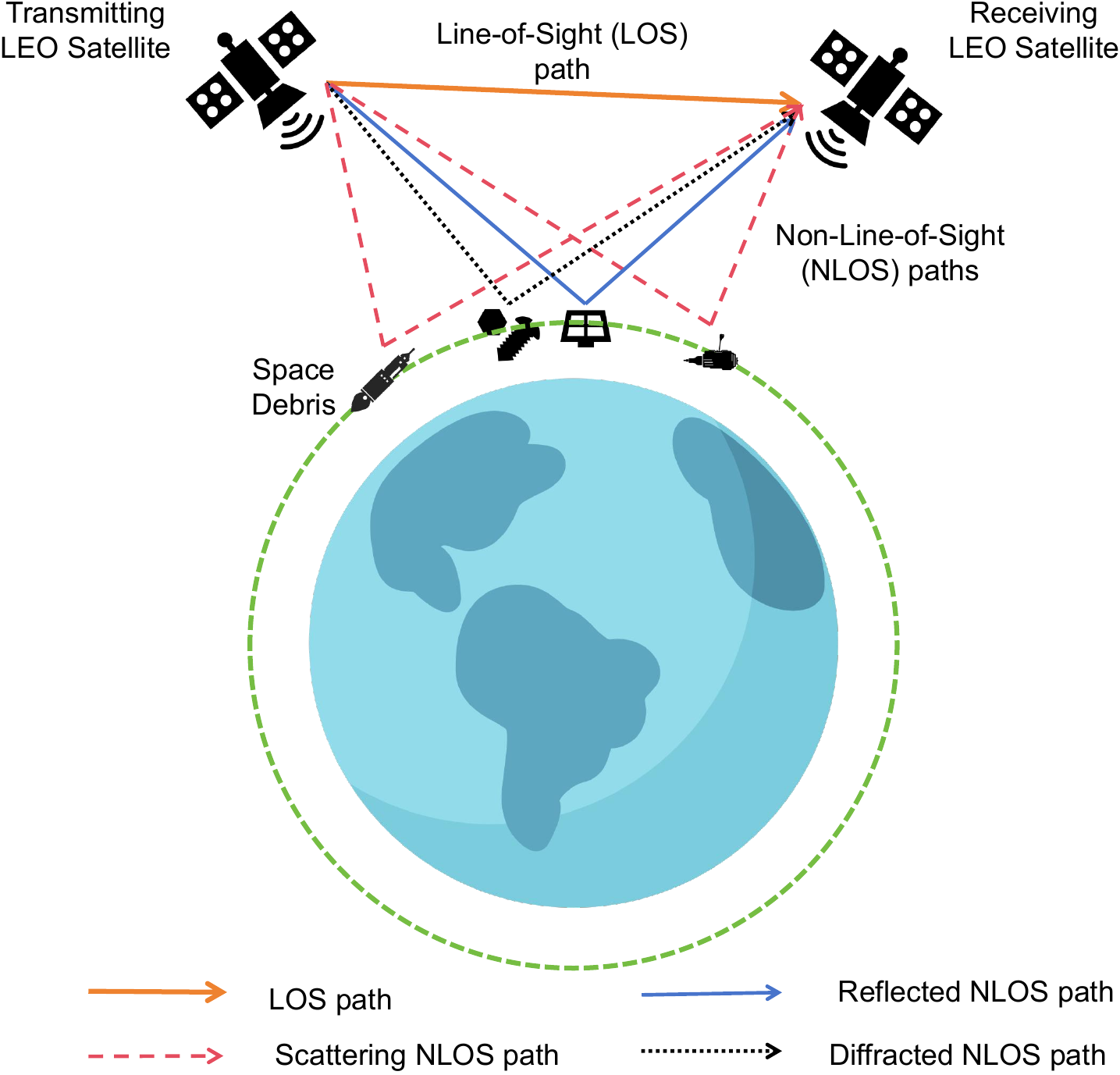}
  \caption{DebriSense-THz scenario illustrating LOS and NLOS signal paths in the presence of space debris.}
  \label{fig:scenario}
\end{figure}

The DebriSense-THz system leverages the unique properties of THz signals for simultaneous communication and sensing in LEO satellite networks. Figure \ref{fig:scenario} illustrates the core concept of our approach.

As shown in Figure \ref{fig:scenario}, the system consists of two main components: a transmitting LEO satellite and a receiving LEO satellite. The transmitting satellite emits THz signals that propagate through space, potentially interacting with debris before reaching the receiving satellite. This interaction forms the basis for our sensing mechanism.

Key elements of the DebriSense-THz system include:

\begin{itemize}
    \item \textbf{Line-of-Sight (LOS) Path:} The direct signal path between the transmitting and receiving satellites, represented by the orange arrow in Figure \ref{fig:scenario}. This path carries the primary communication signal.
    
    \item \textbf{Non-Line-of-Sight (NLOS) Paths:} These are indirect signal paths resulting from interactions with space debris, illustrated by the blue, red, and black dotted lines in Figure \ref{fig:scenario}. NLOS paths include:
    \begin{itemize}
        \item Reflected paths (blue solid line): Signals bouncing off larger debris surfaces.
        \item Scattered paths (red dashed line): Signals dispersed by smaller or irregularly shaped debris.
        \item Diffracted paths (black dotted line): Signals bending around the edges of debris.
    \end{itemize}
    
    \item \textbf{Space Debris:} Various types of orbital debris, including solar panels, rocket components, space asset pieces, represented by different shapes in Figure \ref{fig:scenario}, which interact with the THz signals.
\end{itemize}

The DebriSense-THz system exploits the high sensitivity of THz signals to material properties and physical obstructions. When THz signals encounter space debris, they undergo changes in amplitude, phase, and direction. These changes, captured in the CSI at the receiving satellite, contain valuable information about the presence and characteristics of the debris.

By analysing the received signals, particularly the NLOS components, DebriSense-THz can detect the presence of debris and classify its type. The system employs signal processing techniques and machine learning algorithms, specifically SVMs, to extract relevant features from the received signals and perform debris detection and classification.

This approach allows for continuous sensing of the space environment during routine satellite communications, providing a dual-use functionality that enhances space situational awareness without the need for dedicated sensing hardware.

\subsection{Signal Processing and Feature Extraction}
\begin{figure*}[htbp]
  \centering
  \includegraphics[width=1\textwidth]{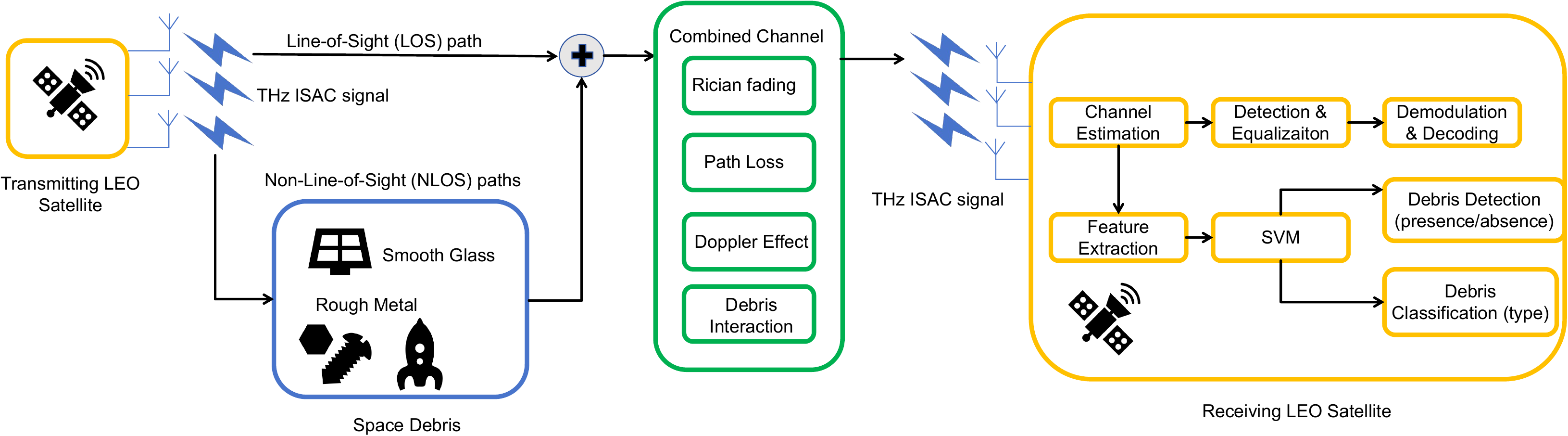}
  \caption{DebriSense-THz system architecture showing signal processing pipeline and debris detection/classification approach.}
  \label{fig:system_architecture}
\end{figure*}

The DebriSense-THz system employs a streamlined signal processing pipeline optimized for on-board implementation in LEO satellites. Figure \ref{fig:system_architecture} illustrates the comprehensive architecture of our proposed system, highlighting the key stages from signal reception to debris classification.

As depicted in Figure \ref{fig:system_architecture}, the signal processing and feature extraction pipeline consists of several key stages:

\begin{enumerate}
    \item \textbf{THz Signal Reception:} The receiving LEO satellite captures the THz signals using a MIMO antenna array. These signals have potentially interacted with space debris, carrying valuable information about the space environment.
    
    \item \textbf{Channel Estimation:} The system performs channel estimation to obtain the CSI. This step is crucial as the CSI encapsulates the effects of the propagation environment, including any interactions with space debris.
    
    \item \textbf{CSI-based Feature Extraction:} Our approach focuses exclusively on CSI-based features due to their high sensitivity to debris interactions and computational efficiency. The feature extraction process involves:
    \begin{itemize}
        \item Calculating the magnitude of the complex CSI.
        \item Computing five key statistical features: mean, variance, maximum, minimum, and skewness of the CSI magnitude.
    \end{itemize}
    These features effectively capture the essential characteristics of the channel affected by the presence and properties of debris.
    
    \item \textbf{SVM-based Debris Detection:} The extracted CSI features are input to a pre-trained SVM model for debris detection. This binary classification determines the presence or absence of debris in the signal path.
    
    \item \textbf{SVM-based Debris Classification:} If debris is detected, a second SVM model is employed to classify the type of debris (e.g., smooth glass, rough metal) based on the same CSI features.
    
    \item \textbf{Alert Triggering:} Based on the detection and classification results, the system triggers appropriate alerts or responses, which can inform satellite operators about potential collision risks or contribute to space situational awareness databases.
\end{enumerate}

The SVM models for both detection and classification are pre-trained on Earth using extensive simulated and real data, capturing a wide range of debris scenarios and THz signal interactions. This approach allows for efficient on-board decision-making without the need for continuous communication with ground stations for debris analysis.

By focusing on CSI-based features and employing a two-stage SVM approach, DebriSense-THz achieves a balance between computational efficiency and detection accuracy. This makes it suitable for real-time implementation on resource-constrained LEO satellites while maintaining high performance in debris detection and classification.

The simplicity of the CSI-based feature set, combined with the power of SVM classification, enables DebriSense-THz to operate as an always-on debris sensing system. It continuously analyzes the space environment during routine satellite communications, providing valuable insights into the evolving debris landscape in LEO without compromising the primary communication functions of the satellite network.

\subsection{SVM-based detection and classification approach}

DebriSense-THz employs SVMs for debris detection and classification, balancing computational efficiency with high accuracy for on-board satellite implementation. Algorithm \ref{alg:debrisense_onboard_updated} outlines the key procedures in our SVM-based approach. The system extracts five statistical features from the CSI magnitude: mean, variance, maximum, minimum, and skewness. These features capture debris-induced signal changes while remaining computationally efficient.

The SVM-based detection and classification process consists of two stages:

\begin{enumerate}
    \item \textbf{Debris Detection:} A binary SVM classifier ($SVM_{detection}$) determines the presence of debris.
    
    \item \textbf{Debris Classification:} If debris is detected, a multi-class SVM classifier ($SVM_{classification}$) categorizes the debris type.
\end{enumerate}

Both SVM models are pre-trained on Earth using simulated and real data, then deployed on-board for real-time processing. This approach enables efficient decision-making without continuous ground station communication.

Key advantages of our SVM-based approach include:

\begin{itemize}
    \item \textbf{Computational Efficiency:} SVMs, once trained, require minimal computational resources for prediction, making them suitable for on-board satellite processing.
    \item \textbf{Robustness:} SVMs perform well even with limited training data and are less prone to overfitting compared to some other machine learning techniques.
    \item \textbf{Adaptability:} The SVM models can be periodically updated with new training data to account for evolving debris populations and environmental conditions.
\end{itemize}

By integrating this approach with the THz ISAC system, DebriSense-THz achieves high-accuracy debris detection and classification while maintaining primary communication functionality, representing a significant advancement in space situational awareness.

\begin{algorithm}
\caption{DebriSense-THz: On-board Signal Processing and Debris Analysis}
\label{alg:debrisense_onboard_updated}
\begin{algorithmic}[1]

\Procedure{ExtractFeatures}{$CSI_{real}$, $CSI_{imag}$}
    \State $CSI_{mag} \gets \sqrt{CSI_{real}^2 + CSI_{imag}^2}$
    \State $features \gets [mean(CSI_{mag}), var(CSI_{mag}),$
    \State \hspace{10mm} $max(CSI_{mag}), min(CSI_{mag}),$
    \State \hspace{10mm} $skew(CSI_{mag})]$
    \State \textbf{return} $features$
\EndProcedure

\Procedure{OnboardProcessing}{$received\_signal$}
    \State $CSI \gets ChannelEstimation(received\_signal)$
    \State $features \gets ExtractFeatures(CSI_{real}, CSI_{imag})$
    \State $debris\_detected \gets Detect(features, SVM_{detection})$
    \If{$debris\_detected$}
        \State $debris\_type \gets Classify(features,$ 
        \State \hspace{13mm} $SVM_{classification})$
        \State TriggerAlert($debris\_type$)
    \EndIf
\EndProcedure

\Procedure{Detect}{$features$, $SVM_{detection}$}
    \State $scaled\_features \gets StandardizeFeatures(features)$
    \State \textbf{return} $SVM_{detection}.predict(scaled\_features)$
\EndProcedure

\Procedure{Classify}{$features$, $SVM_{classification}$}
    \State $scaled\_features \gets StandardizeFeatures(features)$
    \State \textbf{return} $SVM_{classification}.predict(scaled\_features)$
\EndProcedure

\end{algorithmic}
\end{algorithm}

\section{THz Channel Modelling for LEO Satellite Communications}

This section presents a comprehensive multi-ray MIMO channel model for THz communications between LEO satellites, accounting for the dynamic space environment and debris interactions. Our model addresses the complexities of wideband THz propagation, including frequency-selective fading and time-varying characteristics due to rapid satellite and debris motion. We integrate multiple signal interaction mechanisms, including line-of-sight propagation, reflection, scattering, and diffraction, to fully capture the impact of space debris on signal propagation. The model employs Fresnel reflection coefficients, the Beckmann-Kirchhoff scattering model, and Fresnel-Kirchhoff diffraction parameters to accurately represent these interactions across the THz band.

\subsection{Multi-ray THz band Channel Modelling}

The THz channel for LEO satellite communications is characterized by a time-varying MIMO channel matrix $\mathbf{H}(\tau, t)$ of dimensions $N_{t} \times N_{r}$, where $N_{r}$ and $N_{t}$ represent the number of receiving and transmitting antennas, respectively:
\begin{equation}
\mathbf{H}(\tau,t) = \left[h_{mn}(\tau,t)\right]_{m=1, \ldots, N_r ; n=1, \ldots, N_t}
\label{H_1}
\end{equation}
To account for the wideband nature of THz signals, we model the channel as a superposition of narrowband signals in multiple frequency sub-bands. For each transceiver pair, the multipath channel response is:
\begin{equation}
h_{mn}(\tau,t) = \sum_{i=1}^{N_{i}} h_{mn,i}(\tau,t)
\end{equation}
where $N_{i}$ is the number of sub-bands \cite{yang2021analysis}. The response within each sub-band $i$, considering multiple paths ($s$) due to debris, is given by:
\begin{equation}
\begin{aligned}
h_{mn,i}(\tau,t) &= \sum_{s=1}^{N_{s}(t)} a_{mn,i,s}(t) \cdot \delta\left(\tau-\tau_{s}(t)\right) \\
&\quad \cdot D(f,v) \cdot S_{r}\left(\phi_{s}^{A}, \theta_{s}^{A}\right) \cdot S_{t}\left(\phi_{s}^{D}, \theta_{s}^{D}\right)
\end{aligned}
\end{equation}
Here, $a_{mn,i,s}(t)$ is the attenuation of the $s$th path, $\tau_{s}(t) = r_{s}/c$ is the propagation delay, $D(f, v)=e^{-j 2 \pi \cdot f \cdot v/c}$ accounts for the Doppler effect due to relative satellite motion, and $S_{r}$ and $S_{t}$ are steering vectors for the receiver and transmitter, respectively.

The steering vectors, crucial for capturing the spatial characteristics of the THz MIMO channel, are defined as
\begin{equation}
S_{r}\left(\phi_{s}^{A}, \theta_{s}^{A}\right)=\left[1, e^{-j 2 \pi \Delta_{r} \Omega_{rs}}, \ldots, e^{-j 2 \pi(N_{r}-1) \Delta_{r} \Omega_{rs}}\right]^T
\end{equation}
\begin{equation}
S_{t}\left(\phi_{s}^{D}, \theta_{s}^{D}\right)=\left[1, e^{-j 2 \pi \Delta_{t} \Omega_{ts}}, \ldots, e^{-j 2 \pi(N_{t}-1) \Delta_{t} \Omega_{ts}}\right]^T
\end{equation}
where $\Omega_{rs}=\sin(\theta_{s}^{A}) \cos(\phi_{s}^{A}), \Omega_{ts} = \sin(\theta_{s}^{D}) \cos(\phi_{s}^{D})$ are directional cosines, and $\Delta_{r} = d_{r}/\lambda$, $\Delta_{t} = d_{t}/\lambda$ are normalized antenna element spacings.

The unique characteristics of THz propagation in the LEO environment, particularly in the presence of space debris, necessitate a comprehensive multi-ray model. This model integrates direct (LoS), reflected, scattered, and diffracted paths, each crucial for accurate debris detection and characterization. The integrated model for the $i$-th frequency sub-band is expressed as
\begin{equation}
\begin{aligned}
h_{mn,i}(\tau,t) &= a_{mn, \text{LoS}}^{(i)}(t) \cdot \delta(\tau-\tau_{\text{LoS}}(t)) \cdot \\
&I \cdot D(f, v) \cdot S_{r}(\phi_{\text{LoS}}^{A}, \theta_{\text{LoS}}^{A}) \cdot S_{t}(\phi_{\text{LoS}}^{D}, \theta_{\text{LoS}}^{D}) \\
&+ \sum_{p=0}^{N_{\text{Ref}}^{(i)}} a_{mn,\text{Ref}}^{(i, p)}(t) \cdot \delta(\tau-\tau_{\text{Ref}}^{(p)}(t)) \\
&\quad \cdot D(f, v) \cdot S_{r}(\phi_{p}^{A}, \theta_{p}^{A}) \cdot S_{t}(\phi_{p}^{D}, \theta_{p}^{D}) \\
&+ \sum_{q=0}^{N_{\text{Diff}}^{(i)}} a_{mn,\text{Diff}}^{(i, q)}(t) \cdot \delta(\tau-\tau_{\text{Diff}}^{(q)}(t)) \\
&\quad \cdot D(f, v) \cdot S_{r}(\phi_{q}^{A}, \theta_{q}^{A}) \cdot S_{t}(\phi_{q}^{D}, \theta_{q}^{D}) \\
&+ \sum_{u=0}^{N_{\text{Sca}}^{(i)}} a_{mn,\text{Sca}}^{(i, u)}(t) \cdot \delta(\tau-\tau_{\text{Sca}}^{(u)}(t)) \\
&\quad \cdot D(f, v) \cdot S_{r}(\phi_{u}^{A}, \theta_{u}^{A}) \cdot S_{t}(\phi_{u}^{D}, \theta_{u}^{D}),
\end{aligned}
\end{equation}
where $I$ is the LoS indicator (1 if present, 0 otherwise), $a_{mn,*}^{(i,*)}(t)$ are path-specific attenuation factors, $\tau_{*}$ represent propagation delays, $Doppler(f,v)$ accounts for Doppler effects, and $S_{r}$, $S_{t}$ are spatial steering vectors.

This model is particularly significant for THz LEO communications as it captures the complex interactions between the signal and space debris. Each propagation mechanism, reflection, scattering, and diffraction, provides unique information about the characteristics of the debris, which is crucial for our detection and classification algorithms.

Utilizing the Wiener-Khinchin theorem, we relate the complex attenuation factors to their corresponding transfer functions:
\begin{equation}
\left(\begin{array}{c}
a_{\text{LoS}}^{(i)}(t) \\
a_{\text{Ref}}^{(i, p)}(t) \\
a_{\text{Diff}}^{(i, q)}(t) \\
a_{\text{Sca}}^{(i, u)}(t)
\end{array}\right) = \left(\begin{array}{c}
H_{\text{LoS}}(f_{i},t) \\
H_{\text{Ref}}^{(p)}(f_{i},t) \\
H_{\text{Diff}}^{(q)}(f_{i},t) \\
H_{\text{Sca}}^{(u)}(f_{i},t)
\end{array}\right)
\label{eq:transfer_functions}
\end{equation}

These transfer functions integrate the effects of satellite motion and spatial characteristics, providing a comprehensive representation of the THz channel in the presence of space debris. This formulation allows us to isolate and analyze the impact of different types of debris interactions on the received signal, forming the theoretical foundation for our debris detection and classification approach.

\subsection{Line-of-Sight Channel Response}

The environment in which electromagnetic waves propagate between satellites is free space, i.e., a homogeneous medium with a relative permittivity and magnetic permeability of 1 and an electrical permeability of 0. In general, there is only one direct path between two satellites \cite{kaushal2016optical}, which can be represented as 
\begin{equation}
H_{\text{LoS}}(f,t) = H_{\text{FSPL}}^{\text{LoS}}(f,t) \cdot H_{\text{Abs}}(f) \cdot e^{-j 2 \pi f \cdot \tau_{\text{LoS}}(t)}
\end{equation}

In the context of communications between LEO satellites, the molecular absorption effect is considered negligible because of the sparse presence of water and oxygen molecules in low-altitude orbital regions, rendering molecular absorption insignificant for the THz band.

The FSPL, which accounts for the propagation loss, is given by:
\begin{equation}
H_{\text{FSPL}}^{\text{LoS}}(f,t) = \frac{c}{4 \pi \cdot f \cdot d(t)},
\end{equation}
where $c$ denotes the speed of light, $d(t)$ represents the distance between the transmitter and receiver at moment $t$, and $\tau_{\text{LoS}} = \frac{d(t)}{c}$ specifies the arrival time of LoS propagation.

\subsection{Reflected Channel Response}

Reflection from space debris significantly impacts THz signal propagation in LEO environments, as illustrated in Fig. \ref{fig:reflection}. The unique characteristics of THz waves and the diverse nature of space debris necessitate a specialized model for the reflected channel response:

\begin{figure}[htbp]
  \centering
  \includegraphics[width=0.4\textwidth]{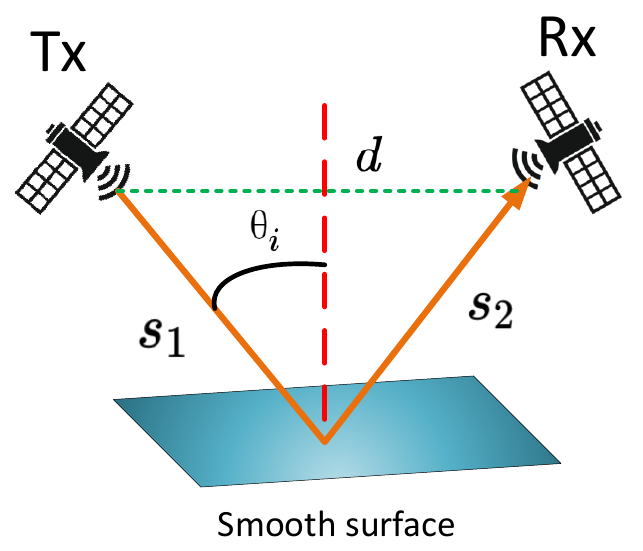}
  \caption{Reflection effect of a THz satellite communication signal via space debris.}
  \label{fig:reflection}
\end{figure}

\begin{equation}
H_{\text{Ref}}(f,t) = H_{\text{FSPL}}^{\text{Ref}}(f,t) \cdot R(f) \cdot e^{-j 2 \pi f \tau_{\text{Ref}}(t)},
\label{eq:reflected_channel}
\end{equation}
where $H_{\text{FSPL}}^{\text{Ref}}(f,t)$ represents the free-space path loss, $R(f)$ is the reflection coefficient, and the exponential term accounts for the phase change due to the additional path length. The delay for the reflected path, $\tau_{\text{Ref}}(t)$, is given by:

\begin{equation}
\tau_{\text{Ref}}(t) = \tau_{\text{LoS}}(t) + \frac{s_{1}(t) + s_{2}(t) - d(t)}{c},
\label{eq:reflected_delay}
\end{equation}
with $s_{1}(t)$ and $s_{2}(t)$ representing the distances from transmitter to reflector and reflector to receiver, respectively, and $d(t)$ the direct path length.

The free-space path loss for the reflected path is expressed as
\begin{equation}
H_{\text{FSPL}}^{\text{Ref}}(f,t) = \frac{c}{4 \pi f (s_{1}(t) + s_{2}(t))}.
\label{eq:reflected_fspl}
\end{equation}

 While Fresnel coefficients typically describe reflection characteristics \cite{piesiewicz2007scattering}, the extremely short wavelengths of THz signals effectively render most debris surfaces rough. A modified reflection coefficient that incorporates a Rayleigh roughness factor has been introduced:
\begin{equation}
R(f) = \rho(f) \cdot \Gamma_{p},
\label{eq:modified_reflection_coeff}
\end{equation}
where $\Gamma_{p}$ represents the Fresnel reflection coefficient for polarization $p$ (TE or TM), and $\rho(f)$ is the roughness coefficient \cite{piesiewicz2007scattering} defined as

\begin{equation}
\rho(f) = e^{-\frac{g}{2}}, \quad g = \left(\frac{4 \pi \sigma \cos \theta_{i}}{\lambda}\right)^{2}
\label{eq:roughness_coeff}
\end{equation}
where $\sigma$ is the standard deviation of the surface height (assumed Gaussian), and $\theta_{i}$ is the incidence angle, calculated based on the geometry shown in Fig. \ref{fig:reflection}:

\begin{equation}
\theta_{i} = \frac{1}{2} \cos^{-1}\left(\frac{s_{1}^{2} + s_{2}^{2} - d^{2}}{2 s_{1} s_{2}}\right).
\label{eq:incidence_angle}
\end{equation}

The Fresnel coefficients $\Gamma_{\text{TE}}$ and $\Gamma_{\text{TM}}$ are calculated using standard formulations \cite{piesiewicz2007scattering}, with the wave impedance of the reflecting material given by:

\begin{equation}
Z = \sqrt{\frac{\mu_{0}}{\varepsilon_{0} \left(n_{ref}^{2} - \left(\frac{\alpha c}{4 \pi f}\right)^{2} - j \frac{2 n \alpha c}{4 \pi f}\right)}},
\label{eq:wave_impedance}
\end{equation}
where $n_{ref}$ is the frequency-dependent refractive index and $\alpha$ is the absorption coefficient, both crucial for accurately modeling various types of space debris in the THz band \cite{taleb2021characterization}.

\subsection{Scattering Channel Response}

\begin{figure}[htbp]
  \centering
  \includegraphics[width=0.4\textwidth]{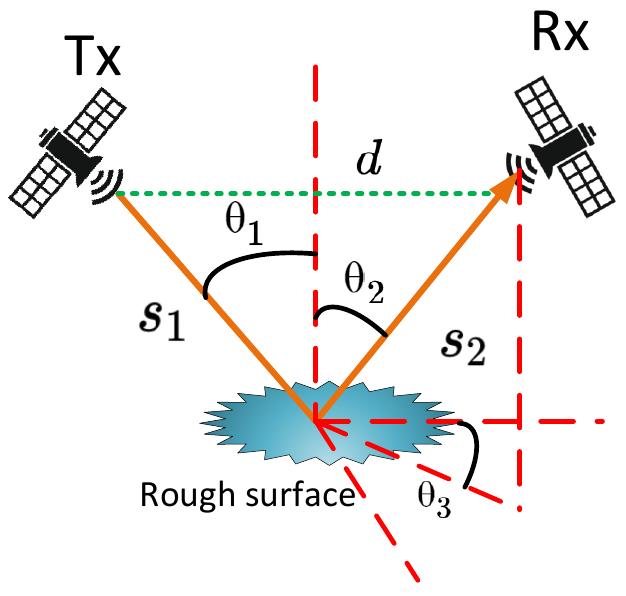}
  \caption{A demonstration of the scattering effect of a satellite communication signal via the debris.}
  \label{fig:scattering}
\end{figure}
Scattering occurs when electromagnetic waves interact with objects or surfaces that are rough relative to the wavelength of the incident signal. In the context of satellite communications, debris can cause significant signal scattering, as shown in Fig. \ref{fig:scattering}. This scattering effect introduces additional multipath components and affects the overall response of the channel. This effect can be expressed as
\begin{equation}
H_{\text{Sca}}(f,t) = H_{\text{FSPL}}^{\text{Sca}}(f,t) \cdot e^{-j 2 \pi f \tau_{\text{Sca}}(t)} \cdot S(f),
\end{equation}
where $H_{\text{FSPL}}^{\text{Sca}}$ represents the free-space path loss specific to scattering, $S(f)$ is the scattering coefficient, and $\tau_{\text{Sca}}$ denotes the time delay associated with the scattering path, calculated as
\begin{equation}
\tau_{\text{Sca}}(t) = \tau_{\text{LoS}} + \frac{s_{1}(t) + s_{2}(t) - d(t)}{c},
\end{equation}
where $s_{1}(t)$ and $s_{2}(t)$ are the distances from the transmitter to the scattering point and from the scattering point to the receiver, respectively, at the moment $t$, while $d(t)$ represents the direct distance between the transmitter and the receiver. The scattering path loss is given by:
\begin{equation}
H_{\text{FSPL}}^{\text{Sca}}(f,t) = \frac{c}{4 \pi f (s_{1}(t) + s_{2}(t))}.
\end{equation}

For the analysis of scattering, we employ the Beckmann-Kirchhoff model  \cite{ragheb2007modified}, which allows us to differentiate scattering coefficients based on wave polarisation. Considering an infinite rectangular area of conductivity $A = l_x l_y$, the scattering coefficient $S(f)$ for an incident wave is determined by the following equation:

\begin{equation}
S_{\text{TE/TM}}(f) = \Gamma_{\text{TE/TM}} \cdot \sqrt{\left(\rho_{0}^{2} + \frac{\pi l_{\text{corr}}^{2} F^{2}}{A} \sum_{m=1}^{\infty} \frac{g_{\text{sca}}^{m}}{m!m} e^{-\frac{v_{\text{xy}}^{2} l_{\text{corr}}^{2}}{4m}}\right) e^{g}},
\end{equation}

where $\Gamma_{\text{TE/TM}}$ represents the Fresnel reflection coefficient for either TE or TM polarization. The incident wave is characterised by the angle of incidence $\theta_{1}$ and is scattered at angles $\theta_{2}$ and $\theta_{3}$. The term $g_{\text{sca}}$ reflects the roughness characteristic of the surface.

It should also be noted that the solution implicitly requires that the lateral dimensions $l_x$ and $l_y$ of the area $A$ be much larger than the wavelength $\lambda$.

The key parameters in this expression include:

\begin{enumerate}
  \item $l_{\text{corr}}$, the correlation length of the surface roughness.
  \item The geometrical factor $F$ derived by Beckmann \cite{beckmann1987scattering} is given by $F=\frac{1+\cos \theta_1 \cos \theta_2-\sin \theta_1 \sin \theta_2 \cos \theta_3}{\cos \theta_2\left(\cos \theta_1 +\cos \theta_2\right)}$.
  \item $v_{x}$ and $v_{y}$, which relate to the spatial frequencies of the surface in the $x$ and $y$ directions, calculated as $v_{x} =k \cdot\left(\sin \left(\theta_{1}\right)-\sin \left(\theta_{2}\right) \cos \left(\theta_{3}\right)\right)$ and $v_{y} = k \cdot (-\sin(\theta_{2}) \sin(\theta_{3}))$ and $v_{\text{xy}} = v_{x}^{2} + v_{y}^{2}$ .
  \item $k$ denotes the free-space wave number.
  \item $\rho_0$ is the magnitude specular reflectance, which can be represented as $\rho_0=\operatorname{sinc}\left(v_x l_x\right) \cdot \operatorname{sinc}\left(v_y l_y\right)$.
  \item $g_{\text{sca}} = k^{2} \sigma^{2} (\cos(\theta_{1}) + \cos(\theta_{2}))^{2}$, representing the roughness factor.
\end{enumerate}

\subsection{Diffracted Channel Response}
\begin{figure}[htbp]
  \centering
  \includegraphics[width=0.4\textwidth]{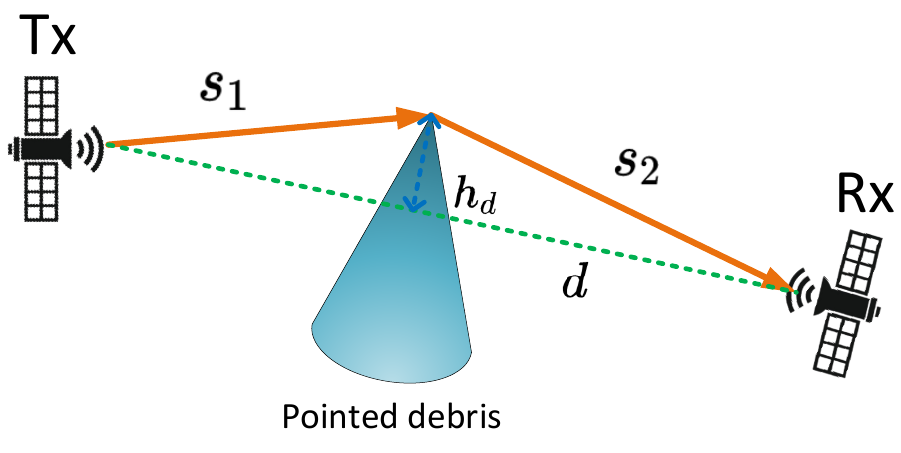}
  \caption{A demonstration of the diffracted effect of a satellite communication signal via the debris.}
  \label{fig:diffraction}
\end{figure}
Diffraction of THz waves around space debris introduces significant complexity to the channel model, particularly when debris size is comparable to or larger than the THz wavelength. This phenomenon, illustrated in Fig. \ref{fig:diffraction}, is crucial for accurately characterizing THz signal propagation in LEO environments.

We model the diffracted channel response as
\begin{equation}
H_{\text{Diff}}(f,t) = H_{\text{FSPL}}^{\text{Diff}}(f,t) \cdot e^{-j 2 \pi f \tau_{\text{Diff}}(t)} \cdot K(f),
\end{equation}
where $H_{\text{FSPL}}^{\text{Diff}}(f,t) = \frac{c}{4 \pi \cdot f \cdot (s_{1}+s_{2})}$ represents the free-space path loss for the diffracted signal, $\tau_{\text{Diff}}(t)$ is the diffraction-induced delay, and $K(f)$ is the frequency-dependent diffraction loss coefficient.

The additional path length due to diffraction is given by:
\begin{equation}
\Delta d = \frac{h_{d}^{2} \cdot (s_{1} + s_{2})}{2 \cdot s_{1} \cdot s_{2}},
\end{equation}
where $h_d$ is the perpendicular distance from the diffraction point to the direct LoS path. The total propagation delay for the diffracted signal can be expressed as
\begin{equation}
\tau_{\text{Diff}}(t) = \tau_{\text{LoS}}(t) + \frac{\Delta d}{c}.
\end{equation}

The diffraction loss $D(f)$ can be empirically determined based on the value of the Fresnel-Kirchhoff diffraction parameter \cite{han2014multi} $v(f)$:
\begin{equation}
D(f) =
\begin{cases}
D_1(f), & 0 < v(f) \leq 1 \\
D_2(f), & 1 < v(f) \leq 2.4 \\
D_3(f), & v(f) > 2.4
\end{cases}
\end{equation}
where:
\begin{equation}
D_1(f) = \mu_1(f) \cdot (0.5 e^{-0.95v(f)}),
\end{equation}
\begin{equation}
D_2(f) = \mu_2(f) \cdot (0.4 - \sqrt{0.12 - (0.38 - 0.1v(f))^2}),
\end{equation}
\begin{equation}
D_3(f) = \mu_3(f) \cdot (\frac{0.225}{v(f)}),
\end{equation}
where $\mu_1(f)$, $\mu_2(f)$, and $\mu_3(f)$ are frequency-dependent parameters chosen to best fit empirical data \cite{han2014multi}.

\section{Performance Evaluation}
This section presents a comprehensive evaluation of the proposed THz-Enabled ISAC Debris Detection and Classification system for LEO satellite networks. We assess the system's performance across a range of operating conditions, focusing on both communication reliability and debris sensing accuracy. Our analysis explores the impact of key system parameters, including the frequency band commonly used for satellite communications (Ka-band) as well as the THz frequency bands (300 GHz to 5 THz), MIMO configurations (4x4 to 64x64), Signal-to-Noise Ratio (SNR) levels (5-25 dB), and varying debris densities. We employ a hybrid channel model that combines Rician fading with reflection, scattering, and diffraction phenomena to simulate realistic space debris scenarios. Through extensive simulations and machine learning-based classification, we investigate the intricate relationships between communication performance and debris detection capabilities. This evaluation not only demonstrates the effectiveness of our proposed system but also provides valuable insights into the trade-offs and optimal operating points for future THz ISAC deployments in space environments.

\subsection{Simulation Setup}
Our simulation framework is designed to evaluate the performance of the THz-Enabled ISAC system across various scenarios and system configurations. We conducted three sets of simulations to investigate the impact of different parameters:

\subsubsection{Debris Density and Frequency Analysis}
\begin{table}[ht]
\centering
\caption{Debris Density and Frequency Simulation Configuration}
\label{tab:debris_density}
\begin{tabular}{|l|l|}
\hline
\textbf{Parameter} & \textbf{Value} \\ \hline
Frequencies (Hz) & $30 \times 10^9$, $300 \times 10^9$, $3 \times 10^{12}$, $5 \times 10^{12}$ \\ \hline
Debris Types & None, Smooth glass, Rough metal \\ \hline
MIMO Size & $16 \times 16$ \\ \hline
Debris Densities (per km$^3$) & $1 \times 10^{-7}$, $5 \times 10^{-7}$, $1 \times 10^{-6}$ \\ \hline
Distance (km) & 500 \\ \hline
Relative Velocity (km/s) & 7 \\ \hline
SNR (dB) & 15 \\ \hline
Modulation & QPSK \\ \hline
\end{tabular}
\end{table}

This simulation set examines the interplay between debris density and signal frequency. We consider four frequency bands spanning from Ka-band to high THz, three debris types, and varying debris densities. The MIMO configuration is fixed at 16x16 to isolate the effects of frequency and debris density.

\subsubsection{Frequency and SNR Analysis}
\begin{table}[ht]
\centering
\caption{Frequency and SNR Simulation Configuration}
\label{tab:freq_snr}
\begin{tabular}{|l|l|}
\hline
\textbf{Parameter} & \textbf{Value} \\ \hline
SNR Values (dB) & 5, 10, 15, 20 \\ \hline
Frequencies (Hz) & $30 \times 10^9$, $3 \times 10^{12}$, $5 \times 10^{12}$ \\ \hline
Debris Types & None, Smooth glass, Rough metal \\ \hline
MIMO Size & $16 \times 16$ \\ \hline
Debris Density (per km$^3$) & $1 \times 10^{-6}$ \\ \hline
Distance (km) & 500 \\ \hline
Relative Velocity (km/s) & 7 \\ \hline
Modulation & QPSK \\ \hline
\end{tabular}
\end{table}

This set of simulations investigates the system performance across different SNR levels and frequencies. We maintain a fixed MIMO configuration and debris density to isolate the effects of SNR and frequency on both communication and sensing performance.

\subsubsection{MIMO Configuration and Frequency Analysis}

\begin{table}[h!]
\centering
\begin{tabular}{|l|l|}
\hline
\textbf{Parameter} & \textbf{Value} \\ \hline
SNR (dB) & 20 \\ \hline
Frequencies (Hz) & $30 \times 10^9$, $300 \times 10^9$, $3 \times 10^{12}$, $5 \times 10^{12}$ \\ \hline
Debris Types & None, Smooth glass, Rough metal \\ \hline
MIMO Configurations & $4 \times 4$, $16 \times 16$, $64 \times 64$ \\ \hline
Debris Density (per km$^3$) & $1 \times 10^{-6}$ \\ \hline
Distance (km) & 500 \\ \hline
Relative Velocity (km/s) & 7 \\ \hline
Modulation & QPSK \\ \hline
\end{tabular}
\caption{Simulation Parameters}
\label{tab:simulation_params}
\end{table}

The final set of simulations explores the impact of MIMO configuration sizes in conjunction with different frequency bands. We maintain a fixed SNR and debris density to focus on how MIMO scaling affects system performance across the frequency spectrum.

For all simulations, we employ a hybrid channel model that combines Rician fading with specific debris interaction effects (reflection, scattering, and diffraction). The channel model is adaptively configured based on the type, density, and frequency of debris considered. Each simulation generates 200 samples to ensure statistical significance.

The debris scenario is modelled as an ellipsoidal volume between the transmitting and receiving satellites, with debris uniformly distributed within this volume. Considering the wavelength of the THz band, we consider scenarios with debris of 1 cm and larger and with a higher density than the current real debris distribution. The probabilities of debris interaction are frequency-dependent and vary based on the type of debris (smooth glass or rough metal).

These comprehensive simulation setups allow us to thoroughly evaluate our THz-Enabled ISAC system's performance under a wide range of realistic space environment conditions.

\subsection{Performance Metrics}
To comprehensively evaluate our THz-Enabled ISAC system for debris detection and classification in LEO satellite networks, we employ metrics that assess both communication reliability and sensing accuracy. These metrics allow us to quantify the system's performance across various operating conditions and parameter configurations.

\subsubsection{Communication Performance Metric}
For assessing the communication performance of our system, we utilize the Bit Error Rate (BER):

\begin{equation}
    \text{BER} = \frac{\text{Number of error bits}}{\text{Total number of transmitted bits}}
    \label{eq:ber}
\end{equation}

BER is a fundamental metric in digital communication systems, directly measuring the reliability of data transmission. It quantifies the proportion of bits that are incorrectly received due to noise, interference, and distortion in the channel. In our LEO satellite network scenario, a lower BER indicates better communication performance, reflecting the system's ability to maintain reliable links in the presence of space debris and other channel impairments.

\subsubsection{Sensing Performance Metrics}
To evaluate the sensing capabilities of our system, we focus on two key aspects: debris detection and debris classification. For both tasks, we use accuracy as our primary metric:

\begin{equation}
    \text{Accuracy} = \frac{\text{Number of correct predictions}}{\text{Total number of predictions}}
    \label{eq:accuracy}
\end{equation}

\paragraph{Detection Accuracy} measures the system's ability to correctly identify the presence or absence of debris. It is calculated as the proportion of correct detections (both true positives and true negatives) out of the total number of detection attempts.

\paragraph{Classification Accuracy} assesses the system's capability to correctly categorize detected debris into predefined types (e.g., 'Smooth glass', 'Rough metal'). It is computed as the ratio of correctly classified debris instances to the total number of classification attempts.

\subsection{Impact of System Parameters on Communication Performance}

In this section, we examine the relationships between various system parameters and the communication performance of our THz-Enabled ISAC system, as measured by BER. Our analysis encompasses the effects of frequency, MIMO configuration, SNR, debris type, and debris density, while also considering advanced aspects of THz communications in the LEO environment.

\subsubsection{Effect of Frequency and MIMO Configuration}
\begin{figure}[htbp]
\centering
\includegraphics[width=\linewidth]{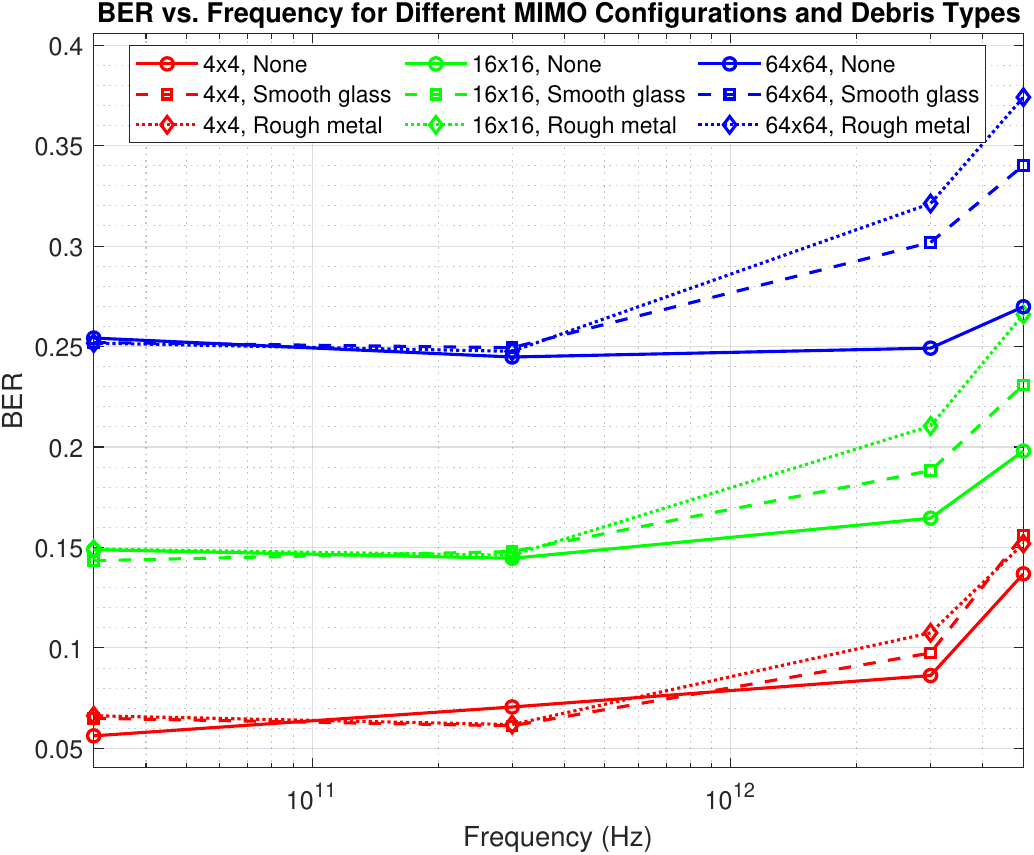}
\caption{BER vs. Frequency for different MIMO configurations and debris types}
\label{fig:BER_vs_Frequency_MIMO_Debris}
\end{figure}

Fig.~\ref{fig:BER_vs_Frequency_MIMO_Debris} illustrates the intricate interplay between frequency, MIMO configuration, and debris types on BER performance. As we transition from lower to higher frequencies, a nuanced picture emerges. In the absence of debris, BER remains relatively stable across the frequency spectrum for all MIMO configurations, demonstrating the system's inherent robustness. However, the introduction of debris, particularly rough metal debris, dramatically alters this landscape.
Larger MIMO configurations, while theoretically offering higher spatial diversity and multiplexing gains, are more susceptible to the Doppler effect and increased phase noise in this dynamic setting. The rapid relative motion between satellites leads to significant Doppler shifts, which become more pronounced across larger antenna arrays. Additionally, the increased complexity of larger MIMO systems makes them more vulnerable to phase synchronisation errors, exacerbating the BER at lower frequencies.

However, the introduction of debris, particularly rough metal debris, dramatically alters this landscape. At 5 THz, we see a marked increase in BER for all MIMO configurations in the presence of debris, with the effect being most pronounced for the 64x64 array. This indicates that while larger MIMO configurations can offer advantages in ideal conditions at higher frequencies, they also become more sensitive to the scattering and reflection effects induced by space debris.

The nature of the debris plays a crucial role in determining the BER performance. Smooth glass debris consistently results in lower BER compared to rough metal debris, especially at higher frequencies. This disparity can be attributed to the different electromagnetic properties of these materials in the THz band, with smooth glass causing less signal scattering and distortion.

\subsubsection{Impact of Debris Density}
\begin{figure}[htbp]
\centering
\includegraphics[width=\linewidth]{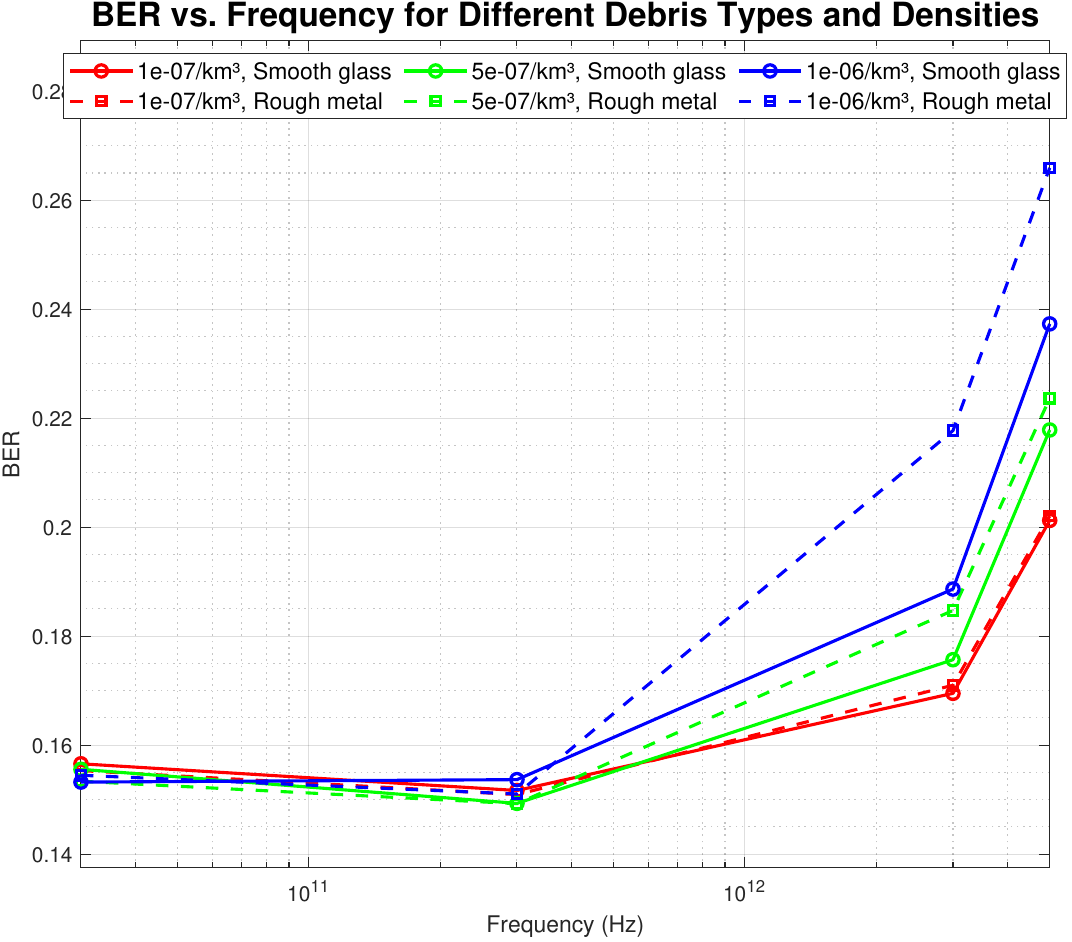}
\caption{BER vs. Frequency for different debris types and densities}
\label{fig:BER_vs_Frequency_Debris_Density}
\end{figure}

The relationship between debris density, frequency, and BER is further elucidated in Fig.~\ref{fig:BER_vs_Frequency_Debris_Density}. As anticipated, increasing debris density correlates with higher BER across all frequencies, a result of heightened signal distortion and multipath effects. However, the magnitude of this impact is not uniform across the frequency spectrum.
In the 3-5 THz range, the effect of debris density on BER becomes particularly pronounced. This frequency-dependent behaviour suggests that as we push into higher THz bands, the system becomes increasingly sensitive to the presence of debris. The compounding effect of density and frequency on BER is evident in the steeper rate of increase in BER with frequency at higher debris densities.

It's important to note that while our simulations explore a range of debris densities to understand system behaviour under various conditions, current LEO environments typically have lower debris densities \cite{esa_space_environment_report}. Recent debris density profiles indicate that the current density of debris larger than 1 cm in LEO is generally below $10^{-6}$ per cubic kilometer. Our simulation scenarios with higher debris densities thus represent potential future conditions or localised areas of higher debris concentration, providing valuable insights into system robustness and potential performance in worst-case scenarios.

\subsubsection{SNR Dependency}
\begin{figure}[htbp]
\centering
\includegraphics[width=\linewidth]{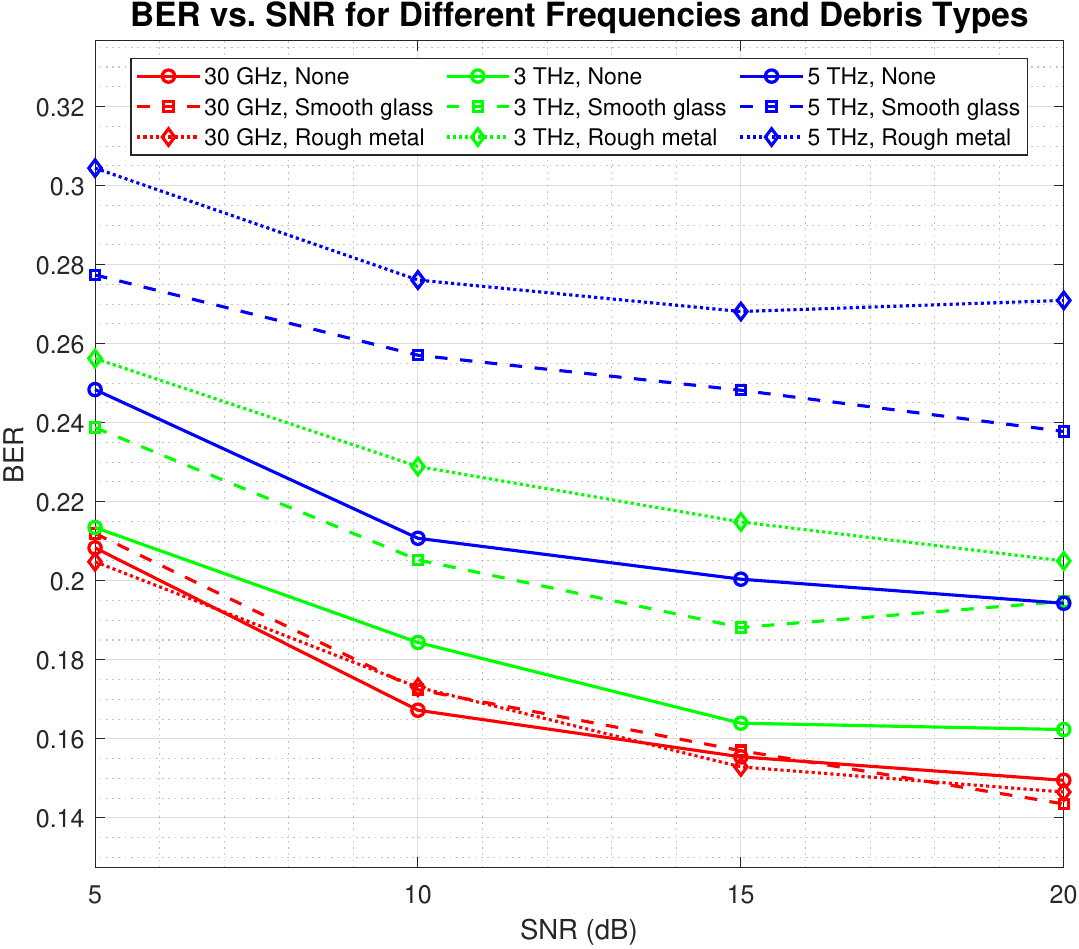}
\caption{BER vs. SNR for different frequencies and debris types}
\label{fig:BER_vs_SNR_Frequency_Debris}
\end{figure}

The relationship between SNR, frequency, and debris type, as depicted in Fig.~\ref{fig:BER_vs_SNR_Frequency_Debris}, reveals more in system behaviour. While the inverse relationship between SNR and BER is preserved across all scenarios, the efficacy of SNR improvements varies significantly with frequency and debris conditions.
At lower frequencies, exemplified by the 30 GHz band, BER shows marked improvement with increasing SNR across all debris scenarios. This behaviour aligns with classical communication theory, where higher signal strength relative to noise typically yields better performance. However, as we transition to higher frequencies, particularly at 5 THz, a different pattern emerges. Here, the BER curves for different debris types converge and show diminished improvement with increasing SNR. This suggests a fundamental shift in the nature of errors at higher frequencies, where debris-induced distortions begin to dominate over noise-induced errors. This observation implies that in debris-rich environments at high THz frequencies, traditional approaches of simply boosting signal power may yield diminishing returns.

\subsection{Debris Detection and Classification Performance}
The efficacy of our THz-Enabled ISAC system in detecting and classifying space debris is crucial for its practical application in LEO environments. This section analyzes the system's sensing performance across various operational parameters, including MIMO configuration, debris density, and SNR levels.

\subsubsection{Impact of MIMO Configuration}
\begin{figure}[htbp]
\centering
\includegraphics[width=1\linewidth]{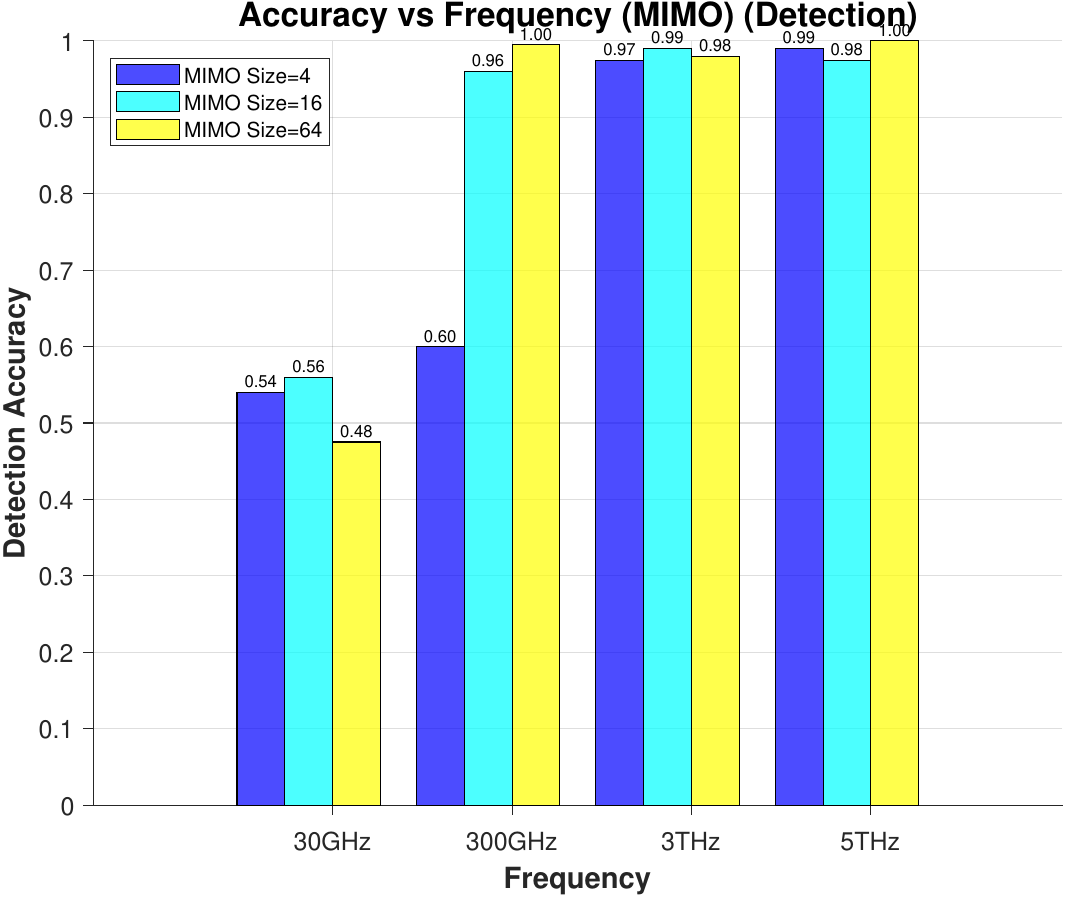}
\includegraphics[width=1\linewidth]{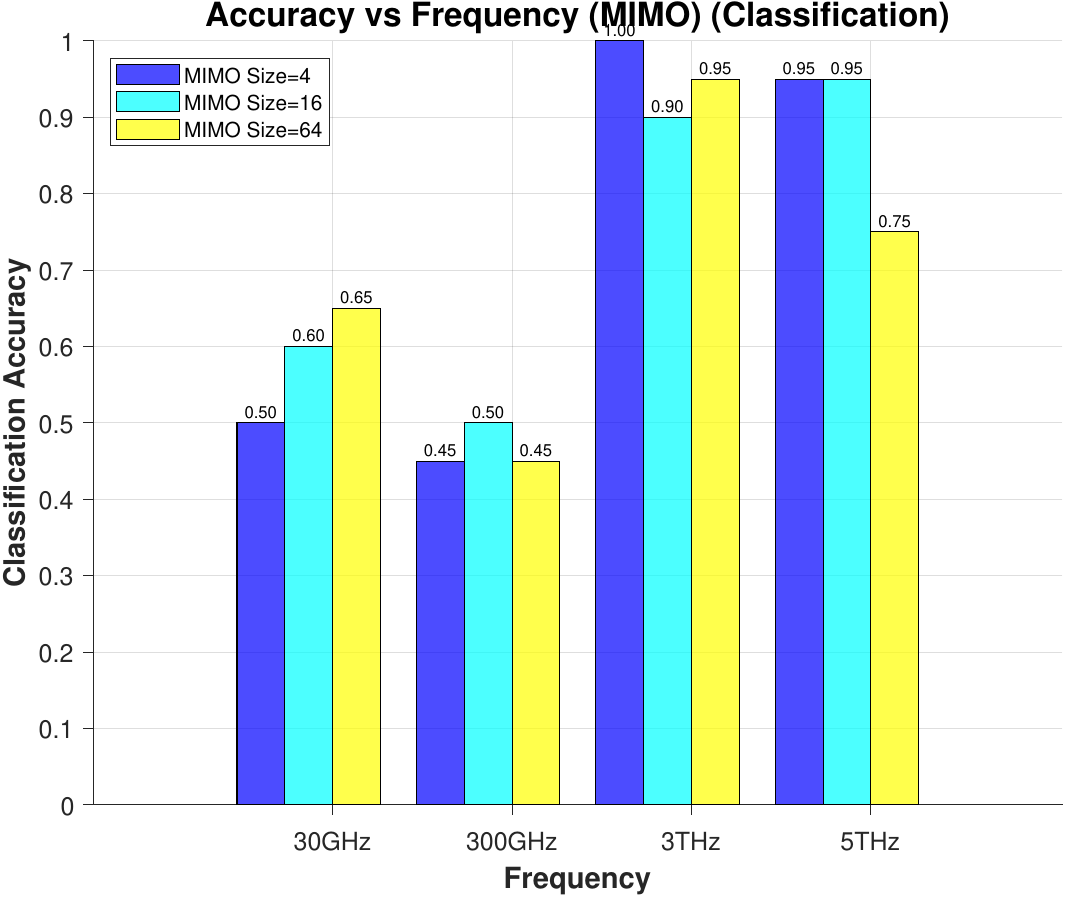}
\caption{Detection (top) and Classification (bottom) Accuracy vs. Frequency for different MIMO configurations}
\label{fig:Accuracy_vs_Frequency_MIMO}
\end{figure}

Fig.~\ref{fig:Accuracy_vs_Frequency_MIMO} illustrates the detection and classification accuracies for different MIMO configurations across the frequency spectrum.

For debris detection, we observe that frequency plays a more dominant role in improving accuracy compared to MIMO configuration size. The performance gap between different MIMO sizes is relatively small, particularly at higher frequencies. At 5 THz, all MIMO configurations achieve high detection accuracy (97-99\%), with the 64x64 setup showing only a marginal improvement over smaller arrays.

Classification accuracy exhibits a similar trend, with frequency being the primary driver of performance improvements. At 30 GHz, larger MIMO configurations show a more pronounced advantage (65\% for 64x64 vs. 50\% for smaller arrays). However, this gap narrows significantly at higher frequencies, with all configurations achieving 95\% accuracy at 3 THz. At 5 THz, we observe a slight divergence, with the 4x4 configuration dropping to 75\% while larger arrays maintain 95\% accuracy.

These results indicate that while larger MIMO configurations offer some advantages in sensing performance, the impact is less significant than initially hypothesized, particularly at higher frequencies. The limited performance gain from increased MIMO size suggests that our current feature extraction method, based on CSI statistical variables, may not fully capture the potential benefits of larger arrays.

It's worth noting that more powerful machine learning techniques, such as Convolutional Neural Networks (CNNs) using CSI matrices as images, could potentially better leverage the spatial information provided by larger MIMO configurations. This approach might reveal more substantial benefits of increased MIMO size, especially in challenging scenarios where subtle spatial variations in the signal become critical for accurate detection and classification.

\subsubsection{Effect of Debris Density}

\begin{figure}[htbp]
\centering
\includegraphics[width=\linewidth]{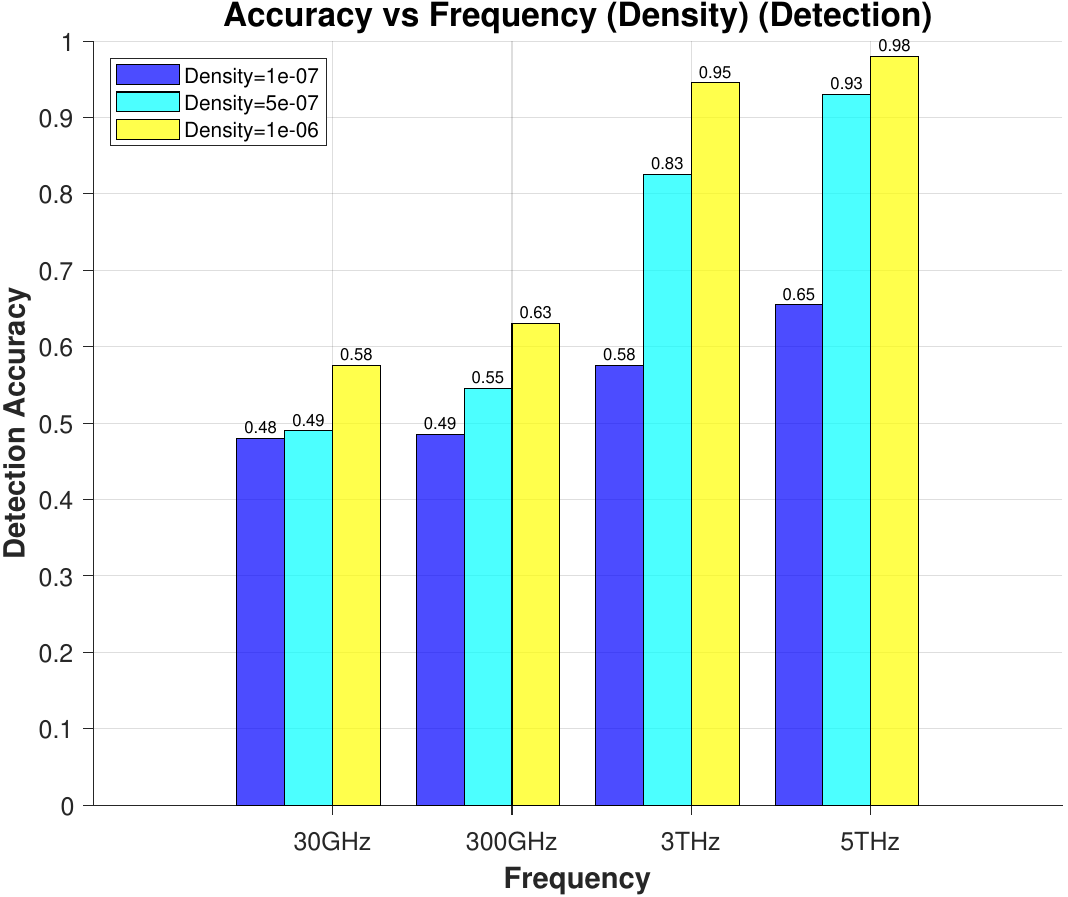}
\includegraphics[width=\linewidth]{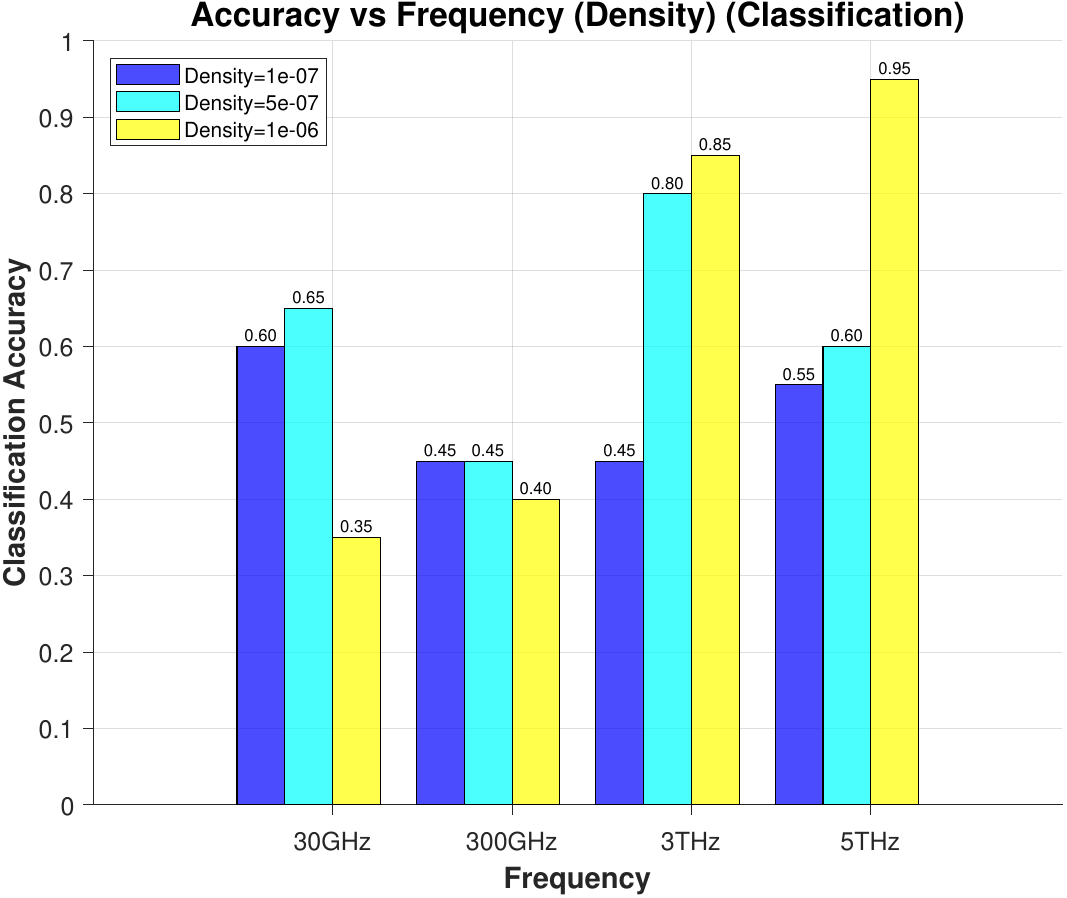}
\caption{Detection (top) and Classification (bottom) Accuracy vs. Frequency for different debris densities}
\label{fig:Accuracy_vs_Frequency_Density}
\end{figure}

The performance of the system under varying debris densities is crucial for understanding its operational capabilities in different LEO environments. Fig.~\ref{fig:Accuracy_vs_Frequency_Density} presents this analysis.

Detection accuracy shows a strong positive correlation with debris density across all frequencies. At 30 GHz, the accuracy increases from 48\% at $10^{-7}$/km³ to 58\% at $10^{-6}$/km³. This trend is amplified at higher frequencies, with 5 THz showing near-perfect detection (98-99\%) for all tested densities.

Classification accuracy exhibits a more complex relationship with debris density. At 30 GHz, higher densities yield better classification (60\% at $10^{-6}$/km³ vs. 45\% at $10^{-7}$/km³). However, this advantage diminishes at 3 THz, where all densities achieve 85-95\% accuracy. 

These results suggest that while higher debris densities generally facilitate better detection, and their impact on classification accuracy is frequency-dependent.

\subsubsection{Influence of SNR}

\begin{figure}[htbp]
\centering
\includegraphics[width=\linewidth]{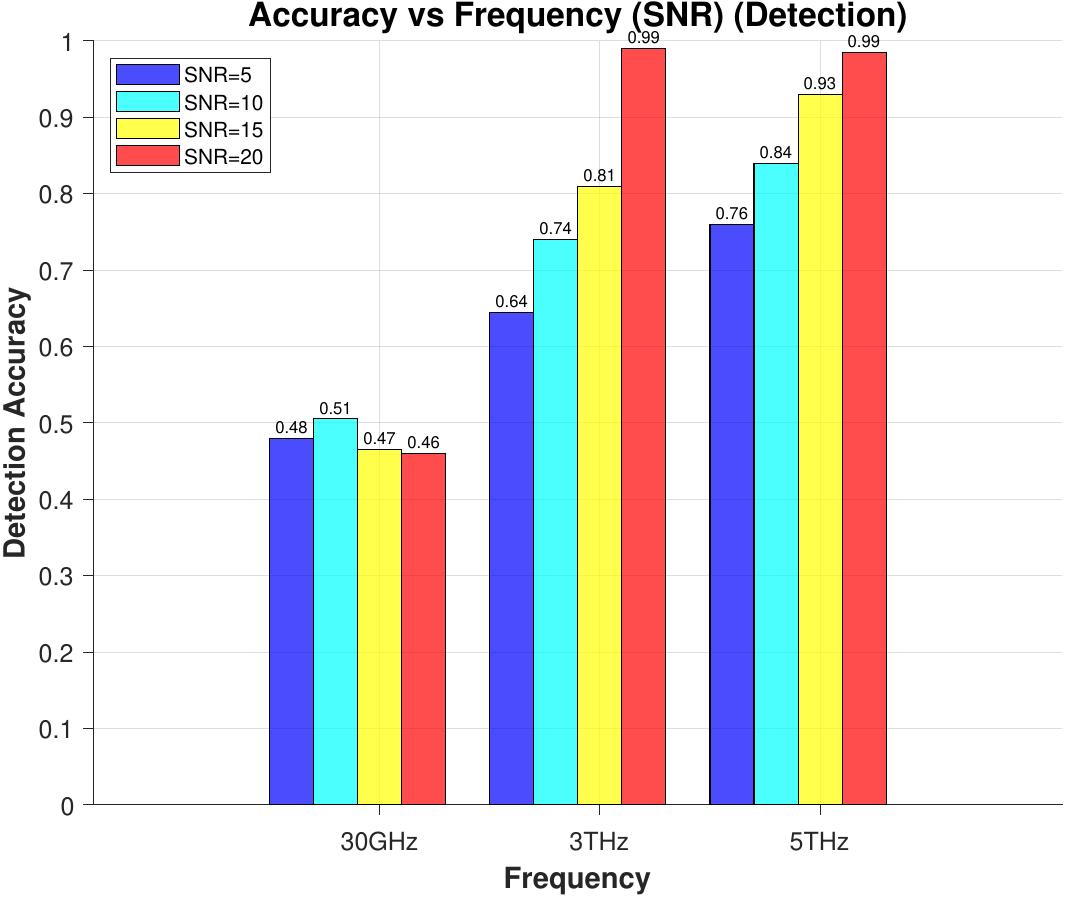}
\includegraphics[width=\linewidth]{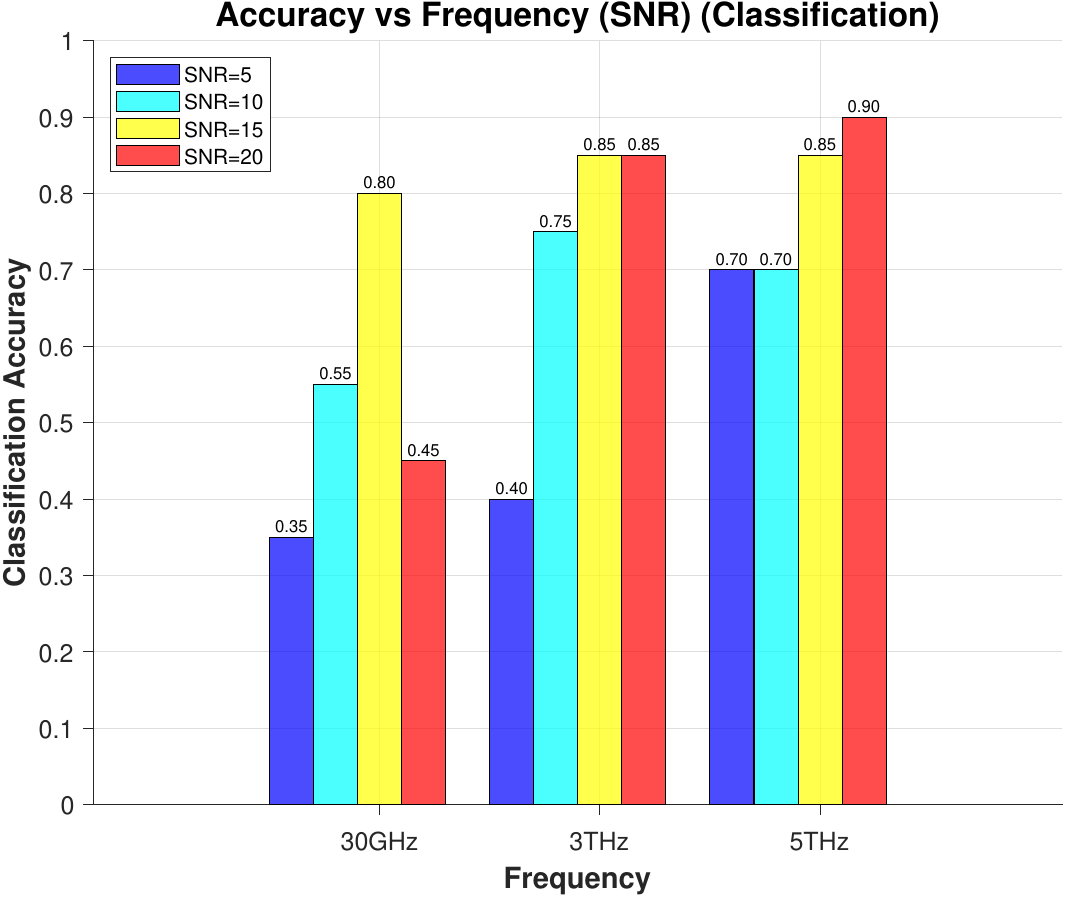}
\caption{Detection (top) and Classification (bottom) Accuracy vs. Frequency for different SNR levels}
\label{fig:Accuracy_vs_Frequency_SNR}
\end{figure}

The system's robustness to varying signal quality is assessed through its performance across different SNR levels, as shown in Fig.~\ref{fig:Accuracy_vs_Frequency_SNR}.

For debris detection, higher SNR levels consistently yield better accuracy across all frequencies. At 30 GHz, accuracy improves from 46\% at 5 dB SNR to 51\% at 20 dB SNR. This improvement is more pronounced at higher frequencies, with 5 THz showing 99\% accuracy for 20 dB SNR compared to 76\% for 5 dB SNR.

Classification accuracy follows a similar trend but with less sensitivity to SNR changes at higher frequencies. At 30 GHz, accuracy ranges from 35\% (5 dB SNR) to 45\% (20 dB SNR). At 5 THz, all SNR levels achieve high accuracy (85-90\%), with only a slight advantage for higher SNR levels.

These findings highlight the system's enhanced robustness to noise at higher frequencies, particularly for classification tasks. This suggests that THz frequencies offer superior debris characterization capabilities even in challenging signal environments.

In conclusion, our THz-Enabled ISAC system demonstrates excellent debris detection and classification capabilities, particularly at higher frequencies. Greater SNR and higher debris densities generally improve performance, while the system shows remarkable resilience to varying MIMO sizes in the THz range. These results underscore the potential of THz ISAC systems for effective space debris sensing and characterisation in LEO environments.

\section{System Performance Trade-offs and Future Directions}

Our overall analysis of the THz-Enabled ISAC system for LEO satellite networks reveals intricate trade-offs between communication and sensing performance, with significant implications for future system designs and operational strategies in space debris sensing and mitigation.

\subsection{Communication-Sensing Performance Balance}
The relationship between communication reliability and sensing accuracy is primarily influenced by operating frequency rather than MIMO configuration. At lower frequencies (30 GHz), a trade-off exists between communication performance and sensing capabilities. However, this trade-off diminishes in the THz range (3-5 THz). While larger MIMO configurations initially show higher BER due to Doppler effects and phase noise in LEO environments, their advantage in debris detection and classification is less pronounced than expected, particularly at THz frequencies. In fact, at 3-5 THz, sensing performance converges across different MIMO sizes, with all configurations achieving high accuracies (95-99\%). These findings suggest that current feature extraction methods may not fully exploit the potential of larger MIMO arrays. Future research should explore advanced machine learning techniques, such as CNNs, to better leverage spatial information in challenging scenarios.

\subsection{Frequency-Dependent Optimization}
Our results indicate that system performance is highly frequency-dependent. While higher frequencies generally offer improved sensing capabilities, they also present challenges for communication reliability, especially in debris-rich environments. The optimal operating frequency appears to be a function of the specific mission requirements and debris environment.

For instance, at 3 THz, we observe a potential "sweet spot" where both communication and sensing performance are balanced. This frequency range offers significant improvements in detection and classification accuracies over lower frequencies, while still maintaining manageable BER levels, especially for larger MIMO configurations.

\subsection{Machine Learning Model Considerations}

The current implementation utilizes SVM with features derived from CSI statistics. Although this approach has shown promising results, there is significant potential for improvement through more powerful machine learning techniques. CNN, for example, could potentially capture the characteristics of the CSI matrix more effectively, leading to improved classification accuracy.

Furthermore, the use of larger MIMO configurations could provide richer features, potentially enhancing the system's sensing capabilities. However, it is crucial to balance these potential improvements against the computational constraints of LEO satellites. Future research should explore lightweight, optimised neural network architectures that can deliver improved performance within the processing limitations of space-based platforms.

\subsection{Adaptive System Design Considerations}

The interplay between system parameters and performance metrics underscores the need for environment-sensitive and adaptive ISAC systems. Future designs should incorporate the following features:

\begin{enumerate}
    \item \textbf{Dynamic MIMO Configuration:} The ability to adjust MIMO array sizes based on the current operational needs, switching between larger configurations for enhanced sensing and smaller arrays for more reliable communication in challenging environments.

    \item \textbf{Frequency Agility:} Systems should be capable of dynamically selecting operating frequencies to optimize for either communication reliability or sensing accuracy based on current mission priorities and debris conditions.

    \item \textbf{Intelligent Power Allocation:} Given the diminishing returns of increasing SNR at higher frequencies, especially for sensing tasks, future systems should employ smart power allocation strategies that balance power consumption with performance requirements.
\end{enumerate}

\subsection{Operational Implications and Future Research Directions}

The performance characteristics of our THz-Enabled ISAC system suggest new operational paradigms for LEO satellite networks:

\begin{enumerate}
    \item \textbf{Dual-Mode Operation:} Satellites could operate in a high-frequency, large-MIMO configuration for routine debris sensing and switch to a more conservative setup for critical communication tasks.

    \item \textbf{Collaborative Sensing:} Given the superior sensing performance at higher frequencies, a subset of satellites in a constellation could be dedicated to debris sensing, sharing this information with other satellites optimized for communication.

    \item \textbf{Integrated Space Situational Awareness:} The debris detection and classification capabilities of the proposed system can be integrated with ground-based radar and other airborne devices to create a more comprehensive space situational awareness network. This multi-layered approach could significantly enhance the accuracy and reliability of debris tracking and characterization.

    \item \textbf{LEO Satellite Constellation Cooperation:} Future research should explore the potential for cooperative ISAC debris detection and classification among LEO satellite constellations. This could involve distributed sensing algorithms that leverage the spatial diversity of multiple satellites to improve overall system performance.

    \item \textbf{3D Debris Trajectory Prediction:} Building upon the high-accuracy debris detection and classification capabilities, future work should focus on establishing 3D debris motion trajectory prediction models. This would enable more proactive collision avoidance strategies and contribute to long-term space sustainability efforts.
\end{enumerate}

These advancements in THz-Enabled ISAC systems offer promising solutions for integrated communication and debris sensing in LEO. Future research should focus on realising these adaptive capabilities, exploring advanced machine learning techniques, and developing cooperative sensing strategies. Such efforts are crucial to address the growing challenges of space debris and to ensure the sustainable use of LEO.

\section{Conclusion}

This paper has presented DebriSense-THz, a novel Terahertz-Enabled Debris Sensing system for LEO satellites that leverages Integrated Sensing and Communication (ISAC) technology. Our work addresses the critical challenge of space debris sensing in increasingly congested LEO environments, making several key contributions to the field.

We developed a comprehensive THz channel model for LEO satellite networks, incorporating complex debris interactions. This model provides a foundation for accurate simulation and analysis of THz signal propagation in debris-rich space environments. The proposed DebriSense-THz architecture demonstrates the feasibility of utilizing THz frequencies for dual-purpose communication and sensing in LEO satellite networks.

Our performance evaluation, conducted across various THz frequencies (0.3--5 THz), MIMO configurations, debris densities, and SNR levels, yielded significant findings:

\begin{enumerate}
    \item THz frequencies (3--5 THz) substantially improve debris detection and classification accuracy (95--99\%) compared to lower bands (62--81\% at 30 GHz), emerging as the dominant factor in performance enhancement.
    \item The impact of MIMO configuration size on sensing performance is less pronounced than initially hypothesized, particularly at higher frequencies. While larger arrays show some advantages, the performance gap between different MIMO sizes narrows significantly in the THz range.
    \item The system demonstrates robust performance across various debris densities, with higher densities generally yielding better accuracy, especially at higher frequencies.
    \item SNR levels show a more notable impact on detection accuracy compared to classification accuracy, particularly in the THz range where the system exhibits resilience to lower SNR conditions.
    \item A trade-off exists between communication reliability and sensing accuracy, especially at lower frequencies, highlighting the need for adaptive system designs in LEO environments.
\end{enumerate}

These results underscore the potential of THz ISAC systems to improve space situational awareness. However, several challenges remain, including the need for advanced machine learning models to better capture CSI characteristics and the development of cooperative sensing strategies among satellite constellations.

Future work should focus on integrating DebriSense-THz with existing space situational awareness systems, exploring 3D debris trajectory prediction, and optimising system designs to balance enhanced sensing capabilities with LEO satellite processing constraints.

In conclusion, DebriSense-THz represents a significant advancement in addressing the space debris challenge. As LEO becomes increasingly congested, such innovative technologies will be crucial to ensuring sustainable near-Earth space utilisation.

\bibliographystyle{ieeetr}  % 根据需要选择引用样式，例如 plain, unsrt, alpha, acm, ieee, 
\bibliography{references}  % ref 是你的BibTeX文件的文件名（不带扩展名）

\begin{IEEEbiographynophoto}{Haofan Dong (hd489@cam.ac.uk) }
is a Ph.D. student in the Internet of Everything (IoE) Group, Department of Engineering, University of Cambridge, UK. He received his MRes in CEPS CDT based in UCL in 2023. His research interests are ISAC, communications in space, THz communications.\\
\end{IEEEbiographynophoto}

\begin{IEEEbiographynophoto}{Ozgur B. Akan (oba21@cam.ac.uk)}
  received his Ph.D. degree from the School of Electrical and Computer Engineering, Georgia Institute of Technology, Atlanta, in 2004. He is currently the Head of the Internet of Everything (IoE) Group, Department of Engineering, University of Cambridge, and the Director of the Centre for NeXt-Generation Communications (CXC), Koc¸ University University. His research interests include wireless, nano-, molecular communications, and the Internet of Everything.\\
\end{IEEEbiographynophoto}

\end{document}